\documentclass[journal,draftcls,onecolumn,12pt,twoside]{IEEEtranTCOM}

\normalsize

\usepackage{mathrsfs}
\usepackage{amsfonts}
\usepackage{amssymb,epic, graphicx}
\usepackage[cmex10]{amsmath}
\usepackage{array}
\usepackage{mdwmath}
\usepackage{mdwtab}
\usepackage{eqparbox}
\usepackage[tight,footnotesize]{subfigure}
\usepackage{ifpdf}

\usepackage{color,xcolor}
\usepackage{graphicx}
\usepackage{amsfonts}
\usepackage{mathrsfs}
\usepackage{amsmath}
\usepackage{amssymb}
\usepackage{enumerate}
\usepackage{color,xcolor}
\usepackage{graphicx}
\usepackage{subfigure}
\usepackage{pifont}
\usepackage[cmex10]{amsmath}

\ifCLASSINFOpdf

\else

\fi

\newtheorem{lemma}{Lemma}

\newtheorem{theorem}{Theorem}
\newtheorem{proposition}{Proposition}
\newtheorem{algorithm}{\textbf{Algorithm}}
\usepackage[square,numbers,sort&compress]{natbib}

\begin{document}

\title{New Techniques for Upper-Bounding the ML Decoding Performance of Binary Linear Codes}

\author{Xiao~Ma,~\IEEEmembership{Member,~IEEE,}
Jia Liu and Baoming Bai,~\IEEEmembership{Member,~IEEE,}
\thanks{This work was presented in part at ISIT'2011.}
\thanks{X.~Ma and J. Liu are with the Department
of Electronics and Communication Engineering, Sun Yat-sen
University, Guangzhou 510275, China. (E-mail: maxiao@mail.sysu.edu.cn)}% <-this % stops a space
\thanks{B.~Bai is with State Key Lab.~of ISN, Xidian University, Xi'an 710071, China. (E-mail: bmbai@mail.xidian.edu.cn)}% <-this % stops a space
 }

\markboth{IEEE Transactions on Communications}%
{Submitted paper}

\maketitle

\begin{abstract}
In this paper, new techniques are presented to either simplify or
improve most existing upper bounds on the maximum-likelihood~(ML)
decoding performance of the binary linear codes over additive white
Gaussian noise~(AWGN) channels. Firstly, the recently proposed union
bound using truncated weight spectrum by Ma~{\em et~al} is
re-derived in a detailed way based on Gallager's first bounding
technique~(GFBT), where the ``good region" is specified by a
sub-optimal list decoding algorithm. The error probability caused by
the bad region can be upper-bounded by the tail-probability of a
binomial distribution, while the error probability caused by the
good region can be upper-bounded by most existing techniques.
Secondly, we propose two techniques to tighten the union bound on
the error probability caused by the good region. The first technique
is based on pair-wise error probabilities. The second technique is
based on triplet-wise error probabilities, which can be
upper-bounded by the fact that any three bipolar vectors form a
non-obtuse triangle. The proposed bounds improve the conventional
union bounds but have a similar complexity since they involve only
the $Q$-function. The proposed bounds can also be adapted to
bit-error probabilities.
\end{abstract}


\IEEEpeerreviewmaketitle

\section{Introduction}
In most scenarios, there do not exist easy ways to compute the exact
decoding error probabilities for specific codes and ensembles.
Therefore, deriving tight analytical bounds is an important research
subject in the field of coding theory and practice. Since the early
1990s, spurred by the successes of the near-capacity-achieving
codes, renewed attentions have been paid to the performance analysis
of the maximum-likelihood~(ML) decoding algorithm. Though the ML
decoding algorithm is prohibitively complex for most practical
codes, tight bounds can be used to predict their performance without
resorting to computer simulations. As shown
in~\cite{Shamai02}\cite{Sason06}, most bounding techniques have
connections to either the 1965 Gallager
bound~\cite{Duman98,Duman98a,Shulman99,Twitto07} or the 1961
Gallager-Fano
bound~\cite{Berlekamp80,Kasami92,Kasami93,Sphere94,TSB94,Divsalar99,Sason00,Zangl01,Divsalar03,Yousefi_04,Yousefi04,Mehrabian06}.
This paper is relevant to the 1961 Gallager-Fano bound, which is
also called Gallager's first bounding technique~(GFBT) in the
literature. Our efforts focus on tightening the simplest
conventional union bound, which is simple but loose and even
diverges in the low-SNR region. Similar to many previously reported
upper bounds surveyed in~\cite{Sason06}, our basic approach is based
on GFBT
\begin{eqnarray}
  {\rm Pr} \{E\} &=& {\rm Pr} \{E,\underline{y}\in\mathcal{R}\} + {\rm Pr}
\{E, \underline{y}\notin \mathcal{R}\} \\
   &\leq& {\rm Pr} \{E,\underline{y}\in\mathcal{R}\} + {\rm Pr}
\{\underline{y}\notin \mathcal{R}\}\label{Gallager_first},
\end{eqnarray}
where $E$ denotes the error event, $\underline{y}$ denotes the
received signal vector, and $\mathcal{R}$ denotes an arbitrary
region around the transmitted signal vector which is usually
interpreted as the ``good region". As pointed out in~\cite{Sason06},
the choice of the region $\mathcal{R}$ is very significant, and
different choices of this region have resulted in various different
improved upper bounds. Intuitively, the more similar the region
$\mathcal{R}$ is to the Voronoi region of the transmitted codeword,
the tighter the upper bound is. However, most existing improved
upper bounds have higher computational complexity than the
conventional union bound.

Different from most of the existing works, we define the good region
using a list decoding algorithm. The basic idea is as
follows. Upper bounds on the word-error probability for the list decoding algorithm~(which is suboptimal) can also be
applied to an ML decoding algorithm, while the list decoding
algorithm can limit competitive candidate codewords.

\textbf{Structure:} The rest of this paper is organized as follows.
In Sec.~\ref{sec2}, we present an upper bound of the angle formed by
any three bipolar vectors, which will be used to upper-bound the
triplet-wise error probabilities. In Sec.~\ref{sec3}, we re-derive,
in a detailed way within the framework of the GFBT, the recently
proposed union bound using truncated weight spectrum by Ma~{\em
et~al}~\cite{Ma10}. On one hand, the truncation technique is helpful
when the whole weight spectrum is unknown or not computable. On the
other hand, the truncation technique can be combined with any other
upper-bounding techniques, potentially resulting in tighter upper
bounds. In Sec.~\ref{sec4}, we propose two techniques to improve the
union bound. The first technique is based on the pair-wise error
probabilities, which can be tightened by employing the independence
of the error event and certain components of the received random
vectors. The second technique is based on the triplet-wise error
probabilities, which is shown to be a non-decreasing function of the
angle formed by the transmitted codeword and the other two
codewords. In Sec.~\ref{sec5}, the proposed bounds are adapted to
ensembles of codes and bit-error probabilities. Numerical examples
are provided in Sec.~\ref{sec6} and we conclude this paper in
Sec.~\ref{sec7}.

\section{Preliminaries}\label{sec2}

\subsection{Geometrical Properties of Binary Codes}
Let $\mathbb{F}_2 = \{0, 1\}$ and $\mathcal{A}_2 = \{-1, +1\}$ be
the binary field and the bipolar signal set, respectively. We use
$W_H(\underline v)$ to denote the Hamming weight of a binary vector
$\underline v \stackrel{\Delta}{=} (v_0, v_1, \cdots, v_{n-1}) \in
\mathbb{F}_2^n$. We use $\|\underline y \|$ to denote the magnitude
of a real vector $\underline y \stackrel{\Delta}{=} (y_0, y_1,
\cdots, y_{n-1}) \in \mathbb{R}^n$, that is, $\|\underline y \| = \sqrt{\sum_{0\leq t < n} y_t^2}$. Let $\mathcal{C}[n,k]$ be a
binary linear block code of dimension $k$ and length $n$ with a
generator matrix $G$ of size ${k\times n}$, that is,
\begin{equation}
    \mathcal{C} \stackrel{\Delta}{=}\left\{{\underline c}\in \mathbb{F}_2^n \mid {\underline c} = {\underline u} G, {\underline u}\in \mathbb{F}_2^k\right\}.
\end{equation}
Let $A_{i,j}$ denote the number of codewords $\underline c =
\underline u G$ with $W_H(\underline u) = i$ and $W_H(\underline c)
= j$. Then $\{A_j \stackrel{\Delta}{=} \sum_{i}A_{i, j}$, $0 \leq j \leq n\}$ is referred
to as the weight spectrum of the given code $\mathcal{C}$.

Consider the binary phase shift keying~(BPSK) mapping $\phi:
\mathbb{F}_2^n \mapsto \mathcal{A}_2^n$ taking $\underline s =
\phi(\underline v)$ by $s_t = 1 - 2v_t$ for $0\leq t \leq n-1$. The
image of $\mathcal{C}$ under this mapping is denoted by
$\mathcal{S}\stackrel{\Delta}{=}\phi(\mathcal{C})$. Hereafter, we
may not distinguish $\underline c \in \mathcal{C}$ from its image
$\underline s \in \mathcal{S}$ when representing a codeword. Let
$d_H(\underline v^{(1)}, \underline v^{(2)}) \stackrel{\Delta}{=}
W_H(\underline v^{(1)} - \underline v^{(2)})$ be the Hamming
distance between two binary vectors $\underline v^{(1)}$ and
$\underline v^{(2)}$. Then their Euclidean distance
$\|\phi(\underline v^{(1)}) - \phi(\underline v^{(2)})\|$ is equal
to $2\sqrt{d_H(\underline v^{(1)}, \underline v^{(2)})}$. Obviously,
the vectors in $\mathcal{A}_2^n$ (hence the bipolar codewords) are
distributed on an $n$-dimensional sphere of radius $\sqrt{n}$
centered at the origin $O$ of $\mathbb{R}^n$. We have the following
lemma.

%-----------------------------------------fig----------------------------------------
\begin{figure}
\centering
  \includegraphics[width=8.5cm]{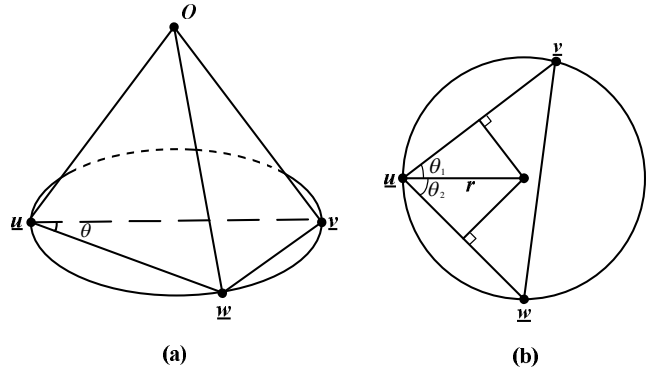}\\
  \caption{Geometrical representation of three bipolar vectors.}\label{Theat_Figure}
\end{figure}
%---------------------------------------------fig-----------------------------------------------------------

\begin{lemma}\label{lemma-upper_bound_of_theta}
Let $\underline u$, $\underline v$ and $\underline w$ be three
bipolar vectors of length $n$. Let $\theta$ be the angle formed by
the two vectors $\overrightarrow{uv} \stackrel{\Delta}{=}\underline
v - \underline u$ and $\overrightarrow{uw}$. Then we have
\begin{equation}
    \theta \leq \min\left\{\frac{\pi}{2},~\arccos\sqrt{\frac{d_1}{n}} + \arccos\sqrt{\frac{d_2}{n}} \right\},
\end{equation}
where $d_1 = d_H(\underline u, \underline v)$ and $d_2 =
d_H(\underline u, \underline w)$.
\end{lemma}
\begin{IEEEproof}
To make the proof more readable, we have drawn the three bipolar
vectors in a three-dimensional space, as shown in
Fig.~\ref{Theat_Figure}~(a). In essence, with a properly chosen
orthogonal transformation, the three vectors can be viewed as three
points in $\mathbb{R}^3$~(a three-dimensional subspace of
$\mathbb{R}^n$). It should be noted that orthogonal transformations
preserve inner products and (hence) lengths as well as angles.

It has been pointed out in~\cite{Agrell98}~(without proof) that any
three bipolar vectors form a non-obtuse triangle, which means
$\theta \leq \pi/2$. For completeness, we re-derive this bound in a
detailed way. Let $\theta$ be the angle formed by
$\overrightarrow{uv}$ and $\overrightarrow{uw}$. It suffices to
prove that the inner product $\overrightarrow{uv} \cdot
\overrightarrow{uw}$ is non-negative. Actually, if $v_t \neq w_t$,
$(v_t-u_t)(w_t-u_t) = 0$ since either $v_t = u_t$ or $w_t = u_t$
must hold; if $v_t = w_t$, $(v_t-u_t)(w_t-u_t) \geq 0$. Therefore
\begin{equation}
    \overrightarrow{uv}\cdot \overrightarrow{uw} = \sum_{t} (v_t-u_t)(w_t-u_t) \geq 0.
\end{equation}

To complete the proof of this lemma, consider the circumscribed
circle of the triangle formed by the three points $\underline u$,
$\underline v$ and $\underline w$~(Fig.~\ref{Theat_Figure}~(b)). Let
$r$ be its radius. The angle can be written as $\theta = \theta_1 +
\theta_2$, where $\cos \theta_1 = \|\overrightarrow{uv}\|/(2r)$ and
$\cos \theta_2 = \|\overrightarrow{uw}\|/(2r)$. It is then not
difficult to verify that
\begin{equation}\label{Angle_Theta}
   \theta = \arccos\frac{\sqrt{d_1}}{r} + \arccos\frac{\sqrt{d_2}}{r}.
\end{equation}
Noticing that the right hand side~(RHS) of~(\ref{Angle_Theta}) is
increasing with $r$ and that $r \leq \sqrt{n}$, we have
\begin{equation}
   \theta \leq  \arccos\sqrt{\frac{d_1}{n}} + \arccos\sqrt{\frac{d_2}{n}}.
\end{equation}
\end{IEEEproof}

\subsection{Union Bounds}

Let $\underline{c}=(c_0, c_1, \cdots, c_{n-1}) \in \mathcal{C}$ be a
codeword. Suppose that $\underline s = \phi(\underline c)$ is
transmitted over an AWGN channel. Let $\underline{y} = {\underline
s} + {\underline z}$ be the received vector, where $\underline z$ is
a vector of independent Gaussian random variables with zero mean and
variance $\sigma^2$. For AWGN channels, the ML decoding is
equivalent to finding the nearest signal vector $\hat{\underline s}
\in \mathcal{S}$ to $\underline y$. A decoding error occurs whenever
$\hat{\underline s} \neq \underline s$. Let $E$ be the decoding
error event~(under ML decoding). Generally, it is a difficult task
to calculate the decoding error probability ${\rm Pr}\{E\}$. Hence
one usually turns to bounding techniques. Due to the symmetry of the channel and the linearity
of the code, the conditional error probability does not depend on the
transmitted codeword, see, e.g.,~\cite{Hof10}. Therefore, without loss of generality, we assume that the all-zero codeword $\underline c^{(0)}$ is transmitted. The
simplest upper bound is the union bound
\begin{eqnarray}\label{error-prob}
  {\rm Pr}\left\{E\right\} &=& {\rm Pr}\left\{\bigcup_{d}E_d\right\}\nonumber \\
    &\leq& \sum_{d} {\rm Pr}\{E_d\}\nonumber\\
    &\leq& \sum_d A_d Q\left(\frac{\sqrt{d}}{\sigma}\right)\label{Union},
\end{eqnarray}
where $E_d$ is the event that there exists at least one codeword of
Hamming weight $d\geq 1$ that is nearer than $\underline{c}^{(0)}$
to $\underline{y}$, and $Q\left(\frac{\sqrt{d}}{\sigma}\right)$ is
the pair-wise error probability with
\begin{equation}\label{Qfunc}
Q(x)\stackrel{\Delta}{=}\int
_{x}^{+\infty}\frac{1}{\sqrt{2\pi}}e^{-\frac{z^{2}}{2}}\,{\rm d}z.
\end{equation}

The question is, how many terms do we need to count for the
summation in the above bound? If too few terms are counted, we will
obtain a lower bound of the upper bound, which may be neither an
upper bound nor a lower bound; if too many are counted, we need pay
more efforts to compute the distance distribution and only a loose
upper bound will be obtained. To get a tight upper bound, we may
determine the terms by analyzing the facets of the Voronoi region of
the codeword $\underline{c}^{(0)}$~\cite{Agrell96}~\cite{Agrell98},
which is a difficult task for a general code.

It is well-known that the conventional union bound is loose and even
diverges~($\geq 1$) in the low-SNR region. One objective of this
paper is, without too much complexity increase, to reduce the number
of involved terms in the conventional union bound. The other
objective of this paper is to tighten the bound on ${\rm
Pr}\{E_d\}$, which used to be upper-bounded by the pair-wise error
probability, where intersections of half-spaces related to codewords
other than the transmitted one are counted more than once. For some
of well-known existing improved bounds based on GFBT, such as the
sphere bound~(SB), the tangential-sphere bound~(TSB) and the
Divsalar bound, see the monograph~\cite[Ch.~3]{Sason06} and the
references therein.

\section{Upper Bounds Using Truncated Weight Spectrum}\label{sec3}

Recently, Ma~{\em et~al}~\cite{Ma10} proposed a union bound which
involves only truncated weight spectrum. In this section, we
re-derive this ``truncated" union bound within the framework of
GFBT, where the region $\mathcal{R}$ is defined in an unusual way
based on the following conceptual suboptimal list decoding
algorithm.

\begin{algorithm}{(A list decoding algorithm for the purpose of
performance analysis)}\label{subDEC}
\begin{enumerate}
  \item[S1.] Make hard decisions, i.e., for $0\leq t \leq n-1$,
\begin{equation}
 \hat{y}_t = \left\{\begin{array}{cc}
                      0, & y_t > 0 \\
                      1, & y_t \leq 0
                    \end{array}
 \right..
\end{equation}
Then the channel $c_t\rightarrow \hat{y}_t$ becomes a memoryless
binary symmetric channel~(BSC) with cross probability $p_{b}
\stackrel{\Delta}{=}Q\left(\frac{1}{\sigma}\right)$.

  \item[S2.] List all codewords within the Hamming sphere with center
at $\underline{\hat{y}}$ of radius $d^*\geq 0$. The resulting list
is denoted as $\mathcal{L}_{\underline{y}}$.
  \item[S3.] If $\mathcal{L}_{\underline y}$ is empty, declare a decoding error; otherwise, find the codeword $\underline{c}^*\in
\mathcal{L}_{\underline{y}}$ such that $\phi(\underline c^*) \in
\mathcal{S}$ is closest to $\underline{y}$.
\end{enumerate}
\hfill\ding{113}
\end{algorithm}

%-----------------------------------------fig-----------------------------------------------------------
\begin{figure}[!t]
\centering
\includegraphics[width = 8.5cm]{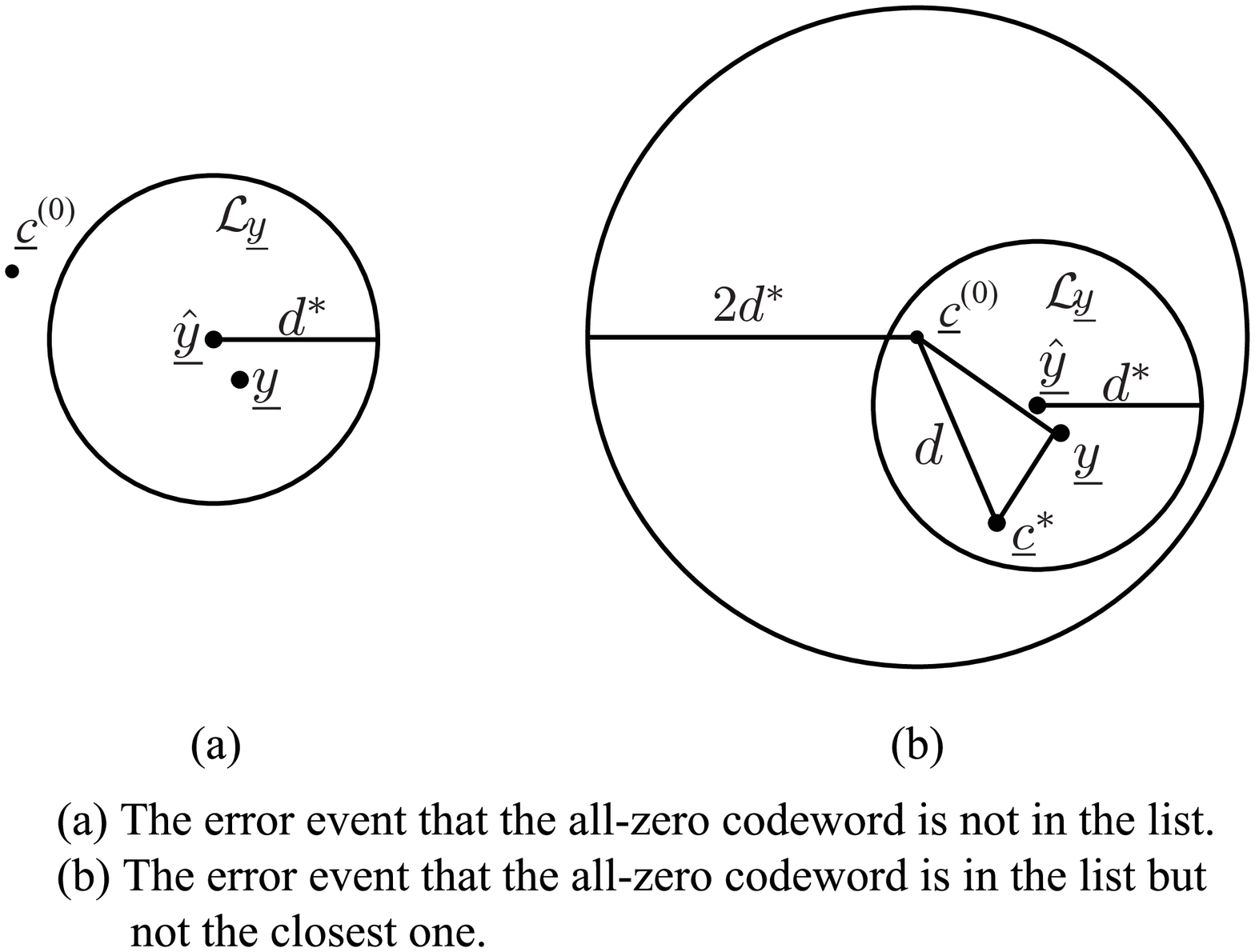}
% where an .eps filename suffix will be assumed under latex,
% and a .pdf suffix will be assumed for pdflatex; or what has been declared
% via \DeclareGraphicsExtensions.
\caption{Graphical illustrations of the decoding error events.}
\label{Error}
\end{figure}
%-----------------------------------------fig-----------------------------------------------------------

Now we define
\begin{equation}\label{Region}
    \mathcal{R} \stackrel{\Delta}{=} \left\{\underline y | {\underline c}^{(0)} \in \mathcal{L}_{\underline y}\right\}.
\end{equation}
In words, the region $\mathcal{R}$ consists of all those $\underline
y$ having at most $d^*$ non-positive components. The decoding error
occurs in two cases under the assumption that the all-zero codeword
${\underline c}^{(0)}$ is transmitted.

{\em Case 1}. The all-zero codeword is not in the list
$\mathcal{L}_{\underline y}$~(see Fig.~\ref{Error}~(a)), that is,
$\underline{y}\notin \mathcal{R}$, which means that at least $d^*
+1$ errors occur over the BSC. This probability is
        \begin{equation}\label{B2}
        {\rm Pr} \{\underline{y}\notin \mathcal{R}\} = \sum_{m=d^*
         +1}^{n}\binom{n}{m}p_{b}^m(1-p_{b})^{n-m}.
        \end{equation}

{\em Case 2}. The all-zero codeword is in the list
$\mathcal{L}_{\underline y}$, but is not the closest one to
$\underline y$~(see Fig.~\ref{Error}~(b)), which is equivalent to
the event $\left\{E, \underline{y} \in \mathcal{R} \right\}$. This
probability is upper-bounded by
        \begin{equation}
         {\rm Pr} \left\{E,\underline{y}\in \mathcal{R} \right\} \leq {\rm Pr} \left\{\bigcup_{d\leq 2d^*} E_d,\;\; \underline{y}\in \mathcal{R} \right\}
        \end{equation}
        since all codewords in the list $\mathcal{L}_{\underline{y}}$ are at
        most $2d^*$ away from the all-zero codeword and not all codewords of a specific weight are in the list. The above upper bound involves only truncated weight spectrum. However, the region $\mathcal{R}$ is in unknown shape and may not be symmetric, which causes difficulties when computing the upper bound. To circumvent this difficulty, we may enlarge $\mathcal{R}$ to $\mathbb{R}^n$ and get
        \begin{eqnarray}\label{B1_Bound}
          {\rm Pr} \left\{E,\underline{y}\in \mathcal{R} \right\}  &\leq& {\rm Pr} \left\{\bigcup_{d\leq 2d^*} E_d,\;\; \underline{y}\in \mathcal{R} \right\}\label{GFBT1a}\\
           &\leq& {\rm Pr} \left\{\bigcup_{d\leq 2d^*} E_d,\;\;\underline{y}\in \mathbb{R}^{n}\right\}\label{GFBT1b}\\
           &=&{\rm Pr}\left\{\bigcup_{d\leq 2d^*} E_d\right\} \leq T_u(\mathcal{C}_{2d^*}), \label{B1}
        \end{eqnarray}
        where $T_u(\mathcal{C}_{2d^*})$ is a computable upper bound on ${\rm Pr}\left\{\bigcup_{d\leq 2d^*} E_d\right\}$, which depends only on the sub-code $\mathcal{C}_{2d^*}$ consisting of all codewords with Hamming weight no greater than $2d^*$. It is worth pointing out that, although the sub-code $\mathcal{C}_{2d^*}$ may not be linear, most bounding techniques in~\cite{Sason06} can be applied to $\mathcal{C}_{2d^*}$ to get such an upper bound under the assumption that the all-zero codeword is transmitted.  Hereafter, we use the notation $\mathcal{C}_{t} \stackrel{\Delta}{=}\left\{{\underline c} \in \mathcal{C} \mid W_H(\underline c) \leq t\right\}$.

For convenience, we define
\begin{equation}\label{Binormial}
    B(p, N_t, N_{\ell}, N_u) \stackrel{\Delta}{=} \sum_{m=N_{\ell}}^{N_u}\binom{N_t}{m}p^m(1-p)^{N_t-m}.
\end{equation}
The function $B(p, N_t, N_{\ell}, N_u)$, which will be used over and
over again in this paper, is just the probability that the number of
bit-errors occurring in a binary vector of total length $N_t$, when
passing through a BSC with cross error probability $p$, ranges from
$N_{\ell}$ to $N_u$. Note that $B(p, N_t, N_{\ell}, N_u)$ can be
calculated recursively independently of codes.

Combining~(\ref{B2}), (\ref{B1}) and (\ref{Binormial}) with
(\ref{Gallager_first}), we get an upper bound
\begin{equation}\label{newbound}
 {\rm Pr}\left\{E\right\} \leq T_u(\mathcal{C}_{2d^*}) +  B(p_b, n, d^*+1, n),
\end{equation}
where the second term in the RHS is computable without requiring the
code structure and the first term depends only on the sub-code
$\mathcal{C}_{2d^*}$.

On one hand, similar to the SB~\cite{Sphere94} and the
TSB~\cite{TSB94}, the proposed upper bound~(\ref{newbound}) involves
only truncated weight spectrum, which is hence helpful when the
whole weight spectrum is not computable. On the other hand, if the
complete weight spectrum is available, the proposed bounding
technique can potentially improve any existing upper bounds.
\begin{proposition}\label{proposition-optnewbound}
Let $T_u$ be an upper-bounding technique. We have
\begin{equation}\label{optnewbound}
 {\rm Pr}\left\{E\right\} \leq \min_{0\leq d^*\leq n} \left\{T_u(\mathcal{C}_{2d^*}) +  B(p_b, n, d^*+1, n)\right\},
\end{equation}
which delivers an upper bound strictly less than 1 and not looser
than any existing upper bounds $T_u(\mathcal{C})$.
\end{proposition}

\begin{IEEEproof}
Noting that $T_u(\mathcal{C}_{0}) = 0 $ and $B(p_b, n, 1, n) =
1-(1-p_b)^n$, we have, by setting $d^* = 0$,
\begin{eqnarray}
{\rm Pr}\left\{E\right\} < 1.
\end{eqnarray}

Similarly, noting that $T_u(\mathcal{C}_{2n}) = T_u(\mathcal{C})$
and $B(p_b, n, n+1, n) = 0$, we have, by setting $d^* = n$,
\begin{eqnarray}
{\rm Pr}\left\{E\right\} \leq T_u(\mathcal{C}).
\end{eqnarray}
\end{IEEEproof}

Taking the conventional union bound as $T_u$, we have
\begin{theorem}\label{theorem-optnewunionbound}
Let $d_{\min}$ be the minimum Hamming weight of the code $\mathcal{C}$. We have
\begin{equation}\label{optnewunionbound}
 {\rm Pr}\left\{E\right\} \leq \min_{0\leq d^*\leq n} \left\{
 \sum_{d_{\min}\leq d\leq 2d^*} A_d Q\left(\frac{\sqrt{d}}{\sigma}\right) +  B(p_b, n, d^*+1, n)\right\}.
\end{equation}
\end{theorem}
\begin{IEEEproof}
It can be proved by substituting the conventional union bound for
$T_u(\mathcal{C}_{2d^*})$~(in the same form as shown
in~(\ref{Union})) into~(\ref{optnewbound}).
\end{IEEEproof}

{\bf Remark.} The bound~(\ref{optnewunionbound}), which is slightly
different from that proposed in~\cite{Ma10}, requires higher
computational loads than the conventional union bound. The overhead
is caused by recursively computing  $B(p_b, n, d^*+1, n)$ and
minimizing over $d^*$. If we do not perform the optimization and
simply set $d^* = n$, we get the conventional union bound, implying
that the technique can potentially improve the conventional union
bound, as stated in Proposition~\ref{proposition-optnewbound}.

\section{Improved Union Bounds}
\label{sec4}

We have interpreted the ``truncated" union bound as an
upper-bounding technique based on the GFBT, where the region
$\mathcal{R}$ is defined by a sub-optimal decoding algorithm. To
bound ${\rm Pr}\{E, {\underline y} \in \mathcal{R}\}$, we have
enlarged $\mathcal{R}$ to $\mathbb{R}^{n}$, as shown in the
derivation from~(\ref{GFBT1a}) to~(\ref{GFBT1b}). The objective of
this section is to reduce the effect of such an enlargement.

Noticing that the event $\underline{y}\in\mathcal{R}$ is equivalent
to the event $W_{H}(\underline{\hat{y}})\leq d^{*}$, we have

\begin{proposition}\label{proposition-optnewunion}
\begin{equation}\label{optnewunion}
  {\rm Pr} \{E\} \leq \min_{0\leq d^*\leq n} \left\{ \sum\limits_{d \leq 2d^{*}}{\rm Pr}\left\{ E_d,
W_{H}(\underline{\hat{y}})\leq d^{*}\right\}+  B(p_b, n, d^*+1,
n)\right\}.
\end{equation}
\end{proposition}
\begin{IEEEproof}
For any $d^*$~($0\leq d^* \leq n$),
\begin{eqnarray}
  {\rm Pr} \{E\} &\leq & {\rm Pr} \{E,\underline{y}\in\mathcal{R}\} + {\rm Pr}
\{\underline{y}\notin \mathcal{R}\}\nonumber \\
   &\leq& {\rm Pr}\left\{\bigcup_{d\leq 2d^*} E_d, W_{H}(\underline{\hat{y}})\leq
  d^{*}\right\} +  B(p_b, n, d^*+1, n) \nonumber\\
  &\leq& \sum\limits_{d \leq 2d^{*}}{\rm Pr}\left\{ E_d,
W_{H}(\underline{\hat{y}})\leq
   d^{*}\right\}+  B(p_b, n, d^*+1, n).
\end{eqnarray}
\end{IEEEproof}

In this section, we focus on how to upper-bound ${\rm Pr}\left\{
E_d, W_{H}(\underline{\hat{y}})\leq d^{*}\right\}$ for any given $d$
and $d^{*}$. Without loss of generality, we assume that $A_d \geq 1$
and denote all the codewords with weight $d$ by $\underline
c^{(\ell)}$, $1\leq \ell\leq A_d$. Let $E_{0\rightarrow \ell}$ be
the event that $\underline c^{(\ell)}$ is nearer than ${\underline
c}^{(0)}$ to $\underline y$.

\subsection{Union Bounds Using Pair-Wise Error Probability}

\begin{lemma}\label{lemma-Improved_Pair_Wise}
\begin{equation}
    {\rm Pr}\left\{ E_{0\rightarrow 1},W_{H}(\underline{\hat{y}})\leq d^{*}\right\} \leq Q(\sqrt{d}/\sigma)B\left(p_b, n-d, 0, d^{*}-1\right).
\end{equation}
\end{lemma}

\begin{IEEEproof}
Without loss of generality, let ${\underline
c}^{(1)}\stackrel{\Delta} {=} (\underbrace{1 \cdots 1}_d
\underbrace{0 \cdots 0}_{n-d}) $. Denote
$\underline{y}_{0}^{d-1}\stackrel{\Delta} {=}(y_0, \cdots, y_{d-1})$
and $\underline{y}_{d}^{n-1}\stackrel{\Delta} {=}(y_d, \cdots,
y_{n-1})$. Evidently, only $\underline{y}_{0}^{d-1}$ can cause the
decoding error event that ${\underline c}^{(1)}$ is nearer than
${\underline c}^{(0)}$ to $\underline{y}$. In other words, the event
$E_{0\rightarrow 1}$ is independent of ${\underline y}_d^{n-1}$ and
${\rm Pr}\{E_{0\rightarrow 1}\} = Q\left(\sqrt{d}/\sigma\right)$.
Also notice that the received signal vector $\underline{y}$ which can cause the event
$E_{0\rightarrow 1}$ must satisfy $W_{H}(\hat{\underline{y}}_0^{d-1}) \geq 1$. Hence $\left\{{\underline y} | E_{0\rightarrow 1},W_{H}(\underline{\hat{y}})\leq d^{*}\right\} \subseteq \left\{{\underline y}| E_{0\rightarrow 1},W_{H}(\hat{\underline y}_{d}^{n-1})\leq d^*-1\right\}$. Then we have
\begin{eqnarray}
  {\rm Pr}\left\{ E_{0\rightarrow 1},W_{H}(\underline{\hat{y}})\leq d^{*}\right\}
   &\leq& {\rm Pr}\left\{ E_{0\rightarrow 1},W_{H}(\hat{\underline y}_{d}^{n-1})\leq d^*-1\right\}\\
   &=& {\rm Pr}\left\{ E_{0\rightarrow 1}\right\}{\rm Pr}\left\{W_{H}(\hat{\underline y}_{d}^{n-1})\leq
   d^*-1\right\}\\
    &=& Q(\sqrt{d}/\sigma)B\left(p_b, n-d, 0, d^{*}-1\right).
\end{eqnarray}

\end{IEEEproof}

\begin{theorem}\label{theorem_ProposedUnion_Pair_Wise}
\begin{equation}\label{ProposedUnion_Pair_Wise}
    {\rm Pr}\left\{ E_d, W_{H}(\underline{\hat{y}})\leq d^{*}\right\} \leq A_{d}Q(\sqrt{d}/\sigma)B\left(p_b, n-d, 0, d^{*}-1\right).
\end{equation}
\end{theorem}

\begin{IEEEproof}
By union bounds and the symmetries of the error events,
\begin{eqnarray}
  {\rm Pr}\left\{ E_d, W_{H}(\underline{\hat{y}})\leq d^{*}\right\} &=&  {\rm Pr}\left\{\bigcup_{1\leq \ell \leq A_d} E_{0\rightarrow \ell},W_{H}(\underline{\hat{y}})\leq d^{*}\right\}\\
  &\leq& \sum\limits_{1\leq \ell \leq A_d} {\rm Pr}\left\{ E_{0\rightarrow \ell},W_{H}(\underline{\hat{y}})\leq d^{*}\right\}\\
   & = & A_{d}{\rm Pr}\left\{ E_{0\rightarrow 1},W_{H}(\underline{\hat{y}})\leq d^{*}\right\}\\
   &\leq & A_{d}Q(\sqrt{d}/\sigma)B\left(p_b, n-d, 0, d^{*}-1\right).
\end{eqnarray}
\end{IEEEproof}

\subsection{Union Bounds Using Triplet-Wise Error Probability}

Temporarily, we assume that $A_d\geq 2$ is even. Then we have
\begin{eqnarray}
  {\rm Pr}\{E_d, W_H(\hat{\underline y}) \leq d^*\}
  &\leq& \sum_{1 \leq \ell \leq A_d/2} {\rm Pr}\left\{E_{0\rightarrow (2\ell - 1)}\bigcup E_{0\rightarrow 2\ell}, W_H(\hat{\underline y}) \leq d^* \right\}.
\end{eqnarray}
If we can find ways to calculate or upper-bound ${\rm
Pr}\left\{E_{0\rightarrow (2\ell-1) }\bigcup E_{0\rightarrow 2\ell},
W_H(\hat{\underline y}) \leq d^*\right\}$, we may improve the
conventional union bound.

In this paper, we refer to the probability ${\rm
Pr}\left\{E_{0\rightarrow 1 }\bigcup E_{0\rightarrow 2}\right\}$ as
{\em triplet-wise error probability}. We have the following lemma.

%-----------------------------------------fig-----------------------------------------------------------
\begin{figure}
\centering
  \includegraphics[width=8.5cm]{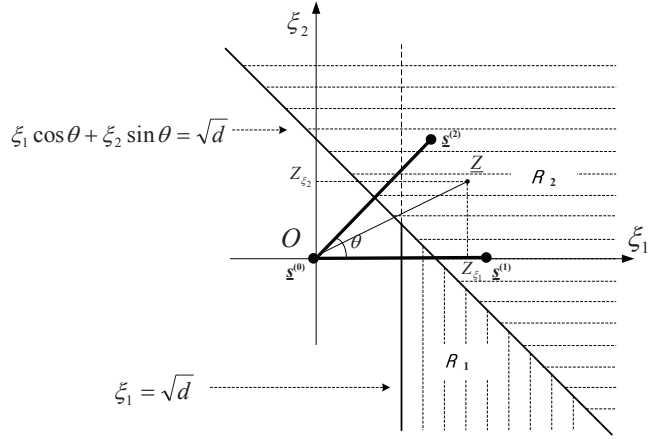}\\
  \caption{Geometrical interpretation of the triplet-wise error probability.}\label{triplet_wise_figure}
\end{figure}
%----
\begin{lemma}\label{lemma-triplet-wise-probability}
Let $\underline c^{(0)}$ be the all-zero codeword with bipolar image
$\underline s^{(0)}$. Let $\underline c^{(1)}$ and $\underline
c^{(2)}$ be the codewords of Hamming weight $d$ with bipolar images
$\underline s^{(1)}$ and $\underline s^{(2)}$, respectively. The
triplet-wise error probability
\begin{equation}\label{R1_R2}
    {\rm Pr}\left\{E_{0\rightarrow 1}\bigcup E_{0\rightarrow 2}\right\}
    = Q(\sqrt{d}/\sigma) +
\int_{\sqrt{d}}^{+\infty}f(\xi_1)\int_{-\infty}^{\frac{\sqrt{d}-\xi_1\cos\theta}{\sin\theta}}f(\xi_2)~{\rm
d}\xi_2~{\rm d}\xi_1,
\end{equation}
where $f(x) = \frac{1}{\sqrt{2\pi}\sigma} e^{-x^2 / (2\sigma^2)}$ is
the probability density function of $\mathcal{N}(0, \sigma^2)$ and
$\theta$ is the angle formed by the two vectors
$\overrightarrow{\underline{s}^{(0)}\underline{s}^{(1)}}$ and
$\overrightarrow{\underline{s}^{(0)}\underline{s}^{(2)}}$.
Furthermore, the triplet-wise error probability is a non-decreasing
function of $\theta$.
\end{lemma}

\begin{IEEEproof}
Similar to the proof of Lemma~\ref{lemma-upper_bound_of_theta}, we
have sketched the two vectors
$\overrightarrow{\underline{s}^{(0)}\underline{s}^{(1)}}$ and
$\overrightarrow{\underline{s}^{(0)}\underline{s}^{(2)}}$ in a
two-dimensional space, as shown in Fig.~\ref{triplet_wise_figure},
where we have chosen $\underline{s}^{(0)}$ as the origin $O$ and
arranged $\overrightarrow{\underline{s}^{(0)}\underline{s}^{(1)}}$
on the abscissa axis $\overrightarrow{O \xi_1}$.

Assume that ${\underline s}^{(0)}$ is transmitted and $\underline y
= {\underline s}^{(0)} + {\underline z}$ is received, where
$\underline z$ is a sample from a random vector $\underline Z$ whose
components are independent and identically distributed as
$\mathcal{N}(0, \sigma^2)$. Let $Z_{\xi_1}$ and $Z_{\xi_2}$ be the
two independent Gaussian random variables by projecting $\underline
Z$ onto the abscissa axis and ordinate axis, respectively.
Specifically, say, $Z_{\xi_1}$ is the inner product $\langle
{\underline Z},
\frac{\underline{s}^{(1)}-\underline{s}^{(0)}}{\|\underline{s}^{(1)}-\underline{s}^{(0)}\|}\rangle$.
It is well-known that only $(Z_{\xi_1}, Z_{\xi_2})$ can cause the
error event $\{E_{0\rightarrow 1}\bigcup E_{0\rightarrow 2}\}$.
Actually,  as shown in Fig.~\ref{triplet_wise_figure}, the error
event $\{E_{0\rightarrow 1}\bigcup E_{0\rightarrow 2}\}$ occurs if
and only if the vector $(Z_{\xi_1}, Z_{\xi_2})$ falls into the
shaded region, which can be partitioned into
\begin{eqnarray}
  \mathcal{R}_1&\stackrel{\Delta}{=}& \left\{(\xi_1,\xi_2)|\xi_1 \geq \sqrt{d}, ~\xi_1\cos\theta + \xi_2\sin\theta< \sqrt{d}\right\},\\
  \mathcal{R}_2&\stackrel{\Delta}{=}& \left\{(\xi_1,\xi_2)|\xi_1\cos\theta + \xi_2\sin\theta \geq \sqrt{d}\right\}.
\end{eqnarray}
Since ${\rm Pr}\{\mathcal{R}_1\}
=\int_{\sqrt{d}}^{+\infty}f(\xi_1)\int_{-\infty}^{\frac{\sqrt{d}-\xi_1\cos\theta}{\sin\theta}}f(\xi_2)~{\rm
d}\xi_2~{\rm d}\xi_1$ and ${\rm Pr}\{\mathcal{R}_2\}
=Q(\sqrt{d}/\sigma) $, we have
\begin{eqnarray}
    {\rm Pr}\left\{E_{0\rightarrow 1}\bigcup E_{0\rightarrow 2}\right\} &=&  {\rm Pr}\{\mathcal{R}_1\} + {\rm Pr}\{\mathcal{R}_2\}\nonumber\\
    &=&\int_{\sqrt{d}}^{+\infty}f(\xi_1)\int_{-\infty}^{\frac{\sqrt{d}-\xi_1\cos\theta}{\sin\theta}}f(\xi_2)~{\rm
d}\xi_2~{\rm d}\xi_1 + Q(\sqrt{d}/\sigma).
\end{eqnarray}

To prove the monotonicity, it suffices to prove that
$\frac{\sqrt{d}-\xi_1\cos\theta}{\sin\theta}$ increases with
$\theta$ for $\xi_1 \geq \sqrt{d}$. This can be verified by noting
that its derivative $\frac{\xi_1-\sqrt{d}\cos\theta}{\sin^{2}\theta}
\geq 0$ for $\xi_1 \geq \sqrt{d}$.
\end{IEEEproof}

\begin{lemma}\label{lemma-UpperTreplet}
For any two codewords ${\underline c}^{(1)}$ and ${\underline
c}^{(2)}$ of Hamming weight $d$, the triplet-wise error
probability\footnote{As pointed out by an anonymous reviewer that
the RHS of~(\ref{UpperTreplet}) is the same as the symbol error
probability of quadrature phase shift keying~(QPSK) over AWGN
channels~\cite{Proakis01}.}
\begin{equation}\label{UpperTreplet}
    {\rm Pr}\left\{E_{0\rightarrow 1}\bigcup E_{0\rightarrow 2}\right\} \leq 2 Q(\sqrt{d}/\sigma) - Q^{2}(\sqrt{d}/\sigma).
\end{equation}
\end{lemma}

\begin{IEEEproof}
From Lemmas~\ref{lemma-upper_bound_of_theta}
and~\ref{lemma-triplet-wise-probability}, we can substitute $\theta
= \pi/2$ into~(\ref{R1_R2}) to complete the proof.
\end{IEEEproof}

{\bf Remark.} From Lemmas~\ref{lemma-upper_bound_of_theta}
and~\ref{lemma-triplet-wise-probability}, in the case of
$\arccos{\sqrt{\frac{d}{n}}} < \pi/4$, we may substitute $\theta =
2\arccos{\sqrt{\frac{d}{n}}}$ into~(\ref{R1_R2}) to get a tighter
bound, however, which needs higher computational loads.

\begin{lemma}\label{lemma-joint-triplet}
For any two codewords ${\underline c}^{(1)}$ and ${\underline
c}^{(2)}$ of Hamming weight $d$,
\begin{equation}\label{joint-triplet}
    {\rm Pr}\left\{E_{0\rightarrow 1}\bigcup E_{0\rightarrow 2}, W_H(\hat{\underline y})\leq d^*\right\} \leq  \left(2 Q(\sqrt{d}/\sigma) - Q^{2}(\sqrt{d}/\sigma)\right) B(p_b, n-2d, 0, d^*-1).
\end{equation}
\end{lemma}
\begin{IEEEproof}
Without loss of generality, assume that
\begin{equation}
{\underline c}^{(1)} \stackrel{\Delta} {=} (c_0^{(1)}\cdots
c_{2d-1}^{(1)} \underbrace{0 \cdots 0}_{n-2d})
\end{equation}
and
\begin{equation}
 {\underline c}^{(2)} \stackrel{\Delta} {=} (c_0^{(2)}\cdots
c_{2d-1}^{(2)} \underbrace{0 \cdots 0}_{n-2d}).
\end{equation}
Then only ${\underline y}_{0}^{2d-1}$ can cause the event that
${\underline c}^{(1)}$ or ${\underline c}^{(2)}$ are nearer than
${\underline c}^{(0)}$ to $\underline{y}$. Also notice that the received signal vector $\underline{y}$ which can cause the event
$E_{0\rightarrow 1}
\bigcup E_{0\rightarrow 2}$ must satisfy $W_{H}(\hat{\underline{y}}_0^{2d-1}) \geq 1$. Hence $\left\{{\underline y}| E_{0\rightarrow 1 }\bigcup E_{0\rightarrow
2},W_{H}(\underline{\hat{y}})\leq d^{*}\right\} \subseteq \left\{{\underline y}| E_{0\rightarrow 1 }\bigcup E_{0\rightarrow
2},W_{H}(\hat{\underline y}_{2d}^{n-1})\leq d^*-1\right\}$. Then we have
\begin{eqnarray}
{\rm Pr}\left\{E_{0\rightarrow 1 }\bigcup E_{0\rightarrow
2},W_H(\underline{\hat{y}})\leq d^{*}\right\}\!\!\!\! &\leq&\!\!\!\!
{\rm Pr}\left\{ E_{0\rightarrow 1}\bigcup E_{0\rightarrow
2},W_H(\hat{\underline y}_{2d}^{n-1})\leq
d^{*}-1\right\}\\
\!\!\!\!&=&\!\!\!\!{\rm Pr}\left\{E_{0\rightarrow 1}\bigcup
E_{0\rightarrow 2}\right\}{\rm Pr}\left\{W_H(\hat{\underline
y}_{2d}^{n-1})\leq
d^{*}-1\right\}\\
\!\!\!\!&\leq&\!\!\!\!\left(2 Q(\sqrt{d}/\sigma) -
Q^{2}(\sqrt{d}/\sigma)\right)B(p_b, n-2d, 0, d^*-1)
\end{eqnarray}
from Lemma~\ref{lemma-UpperTreplet}.
\end{IEEEproof}

The main result of this subsection is the following theorem, which
shows that the union bound based on triplet-wise error probabilities
can be tighter than the conventional union bound based on pair-wise
error probabilities.

\begin{theorem}\label{ProposedUnion_Triplet_Wise}
If $A_d$ is even,
\begin{equation}
  {\rm Pr}\{E_d, W_H(\hat{\underline y}) \leq d^*\} \leq A_{d}\left(
  Q(\sqrt{d}/\sigma)-\frac{1}{2}Q^{2}(\sqrt{d}/\sigma)\right) B(p_b, n-2d, 0, d^*-1);\label{EvenBound}
\end{equation}
if $A_d$ is odd,
\begin{eqnarray}
  {\rm Pr}\{E_d, W_H(\hat{\underline y}) \leq d^*\} &\leq& (A_{d}-1)\left(Q(\sqrt{d}/\sigma)-\frac{1}{2}Q^{2}(\sqrt{d}/\sigma)\right)
   B(p_b, n-2d, 0, d^*-1)\nonumber\\
    &&+
   Q(\sqrt{d}/\sigma)B(p_b, n-d, 0, d^*-1).\label{OddBound}
\end{eqnarray}

\end{theorem}

\begin{IEEEproof}
If $A_d$ is even, we have
\begin{eqnarray}
  {\rm Pr}\{E_d, W_H(\hat{\underline y}) \leq d^*\}
  &\leq& \sum_{1 \leq \ell \leq A_d/2} {\rm Pr}\left\{E_{0\rightarrow (2\ell - 1)}\bigcup E_{0\rightarrow 2\ell}, W_H(\hat{\underline y}) \leq d^* \right\}\nonumber\\
  &\leq& \frac{A_d}{2}\left(2 Q(\sqrt{d}/\sigma) - Q^{2}(\sqrt{d}/\sigma)\right) B(p_b, n-2d, 0, d^*-1)\nonumber\\
  &=&A_{d}\left(
  Q(\sqrt{d}/\sigma)-\frac{1}{2}Q^{2}(\sqrt{d}/\sigma)\right) B(p_b, n-2d, 0, d^*-1),
\end{eqnarray}
which follows from the symmetries of the error events and
Lemma~\ref{lemma-joint-triplet}.

If $A_d$ is odd, we have
\begin{eqnarray}
 \lefteqn{{\rm Pr}\{E_d, W_H(\hat{\underline y}) \leq d^*\} }\nonumber \\
 &\leq& \sum_{1 \leq \ell \leq (A_d-1)/2}{\rm Pr}\left\{E_{0\rightarrow (2\ell-1)}
\bigcup E_{0\rightarrow 2\ell}, W_H(\hat{\underline y}) \leq d^*\right\} + {\rm Pr}\left\{E_{0\rightarrow A_d}, W_H(\hat{\underline y}) \leq d^{*}\right\}\\
  &\leq& (A_{d}-1)\left(Q(\sqrt{d}/\sigma)-\frac{1}{2}Q^{2}(\sqrt{d}/\sigma)\right)
   B(p_b, n-2d, 0, d^*-1)\nonumber\\
   && +Q(\sqrt{d}/\sigma)B(p_b, n-d, 0, d^*-1),
\end{eqnarray}
which follows from the symmetries of the error events and
Lemmas~\ref{lemma-Improved_Pair_Wise} and~\ref{lemma-joint-triplet}.
\end{IEEEproof}

Note that the bounds in Theorem~\ref{ProposedUnion_Triplet_Wise}
will not always improve the bounds in
Theorem~\ref{theorem_ProposedUnion_Pair_Wise}, since it may happen
that $B(p_b, n-2d, 0, d^*-1) > B(p_b, n-d, 0, d^*-1)$.

\section{Adaptations of the Improved Union Bounds}\label{sec5}

\subsection{Bounds for An Ensemble of Codes}
As we know, most existing bounds are applied to ensembles of codes
as well as specific codes. However, the bounds given in
Theorem~\ref{ProposedUnion_Triplet_Wise} can not be applied directly
to ensembles of codes because the average weight spectra of a code
ensemble are usually not be integer-valued.
\begin{theorem}\label{Theorem_UnifiedBound}
Consider a code ensemble $\mathscr{C}$ with probability distribution
${\rm Pr}\{\mathcal{C}\}$, $\mathcal{C} \in \mathscr{C}$. Let
$\{A_d^\mathcal{C}\}$ be the weight spectrum of a specific code
$\mathcal{C}$. Then $A_d = \sum_{\mathcal{C}}{\rm Pr}\{\mathcal{C}\}
A_d^\mathcal{C}$ is referred to as the average weight spectra.
Define

\begin{eqnarray}
    h(A_d) \stackrel{\Delta}{=}\min \left\{
    \begin{array}{c}
       A_d Q(\sqrt{d}/\sigma)B(p_b, n-d, 0, d^*-1), \\
       (A_{d}-1)\left(Q(\sqrt{d}/\sigma)-\frac{1}{2}Q^{2}(\sqrt{d}/\sigma)\right)B(p_b, n-2d, 0, d^*-1)+
       Q(\sqrt{d}/\sigma)
    \end{array}
    \right\}.
\end{eqnarray}

Then ${\rm Pr}\{E_d, W_H(\hat{\underline y}) \leq d^*\} \leq
h(A_d)$.
\end{theorem}

\begin{IEEEproof}
From Theorem~\ref{theorem_ProposedUnion_Pair_Wise}, we have
\begin{eqnarray}
 {\rm Pr}\{E_d, W_H(\hat{\underline y}) \leq d^*\}&=& \sum_{\mathcal{C}}{\rm Pr}\{\mathcal{C}\} {\rm Pr}\{E_d,W_H(\hat{\underline y}) \leq d^*|\mathcal{C}\}\nonumber \\
                   &\leq& \sum_{\mathcal{C}}{\rm Pr}\{\mathcal{C}\}A_d^{\mathcal{C}}Q(\sqrt{d}/\sigma)B(p_b, n-d, 0, d^*-1)\nonumber\\
                   &=&A_d Q(\sqrt{d}/\sigma)B(p_b, n-d, 0, d^*-1).\label{pairUb}
\end{eqnarray}

It can be verified from Theorem~\ref{ProposedUnion_Triplet_Wise}
that, for any $A_d^\mathcal{C}\geq 0$,
\begin{eqnarray}
  {\rm Pr}\{E_d, W_H(\hat{\underline y}) \leq d^*|\mathcal{C}\} &\leq& (A_{d}^\mathcal{C}-1)\left(Q(\sqrt{d}/\sigma)-\frac{1}{2}Q^{2}(\sqrt{d}/\sigma)\right)
   B(p_b, n-2d, 0, d^*-1)\nonumber\\
    &&+
   Q(\sqrt{d}/\sigma).\label{UnifiedBound}
\end{eqnarray}
Then, we have
\begin{eqnarray}
   \lefteqn{ {\rm Pr}\{E_d, W_H(\hat{\underline y}) \leq d^*\}}\nonumber\\
    &  \!\!\!\!\!\!=& \sum_{\mathcal{C}}{\rm Pr}\{\mathcal{C}\} {\rm Pr}\{E_d,W_H(\hat{\underline y}) \leq d^*|\mathcal{C}\}\nonumber \\
    & \!\!\!\!\!\!\leq& \!\!\!\!\sum_{\mathcal{C}}{\rm Pr}\{\mathcal{C}\}\!\!\left\{\!\!(A_d^{\mathcal{C}}-1)\!\!\left(Q(\sqrt{d}/\sigma)-\frac{1}{2}Q^{2}(\sqrt{d}/\sigma)\right)\!\!B(p_b, n-2d, 0, d^*-1)+
       Q(\sqrt{d}/\sigma)\right\}\nonumber\\
    & \!\!\!\!\!\!=&(A_{d}-1)\left(Q(\sqrt{d}/\sigma)-\frac{1}{2}Q^{2}(\sqrt{d}/\sigma)\right)B(p_b, n-2d, 0, d^*-1)+
       Q(\sqrt{d}/\sigma).\label{tripletUb}
\end{eqnarray}
Combining~(\ref{pairUb})~and~(\ref{tripletUb}), and taking into
account the definition of $h(A_d)$, we have
\begin{equation}\label{RefUb}
    {\rm Pr}\{E_d, W_H(\hat{\underline y}) \leq d^*\} \leq h(A_d).
\end{equation}

\end{IEEEproof}

We now summarize the main result in the following theorem, which can
be applied to both specific codes and ensembles of codes.

\begin{theorem}\label{theorm_TotalBound}
Let $\left\{A_{d}\right\}$ be the (average) weight spectrum of a
specific code or a code ensemble. The word-error probability can be
upper-bounded by
\begin{equation}\label{TotalBound}
    {\rm Pr} \{E\} \leq \min_{0 \leq d^* \leq n}\left\{\sum\limits_{ d \leq 2d^*} h(A_d) + B(p_b, n, d^*+1, n)\right\}.
\end{equation}
\end{theorem}

\begin{IEEEproof}
Since a specific code is a special case of a code ensemble with a
degraded probability distribution, we consider only a code ensemble.

Combining Theorem~\ref{Theorem_UnifiedBound} with
Proposition~\ref{proposition-optnewunion}, or equivalently,
substituting~(\ref{RefUb}) into~(\ref{optnewunion}), we then
have~(\ref{TotalBound}), completing the proof.
\end{IEEEproof}

\subsection{Bounds for Bit-Error Probabilities}
In order to adapt the upper bound~(\ref{TotalBound}) to the
bit-error probability, we define
\begin{eqnarray}
  \hat{i}_d &\stackrel{\Delta}{=}& \max\left\{i\mid A_{i, d}>
    0\right\},  \\
     A'_d &\stackrel{\Delta}{=}& \sum_{i}\frac{i}{k}A_{i, d}
\end{eqnarray}

and

\begin{eqnarray}
   h'(A_d) \stackrel{\Delta}{=}\min\left\{\!\!
    \begin{array}{c}
       A'_d Q(\sqrt{d}/\sigma)B(p_b, n-d, 0, d^*-1), \\
       \frac{\hat{i}_d}{k}\left((A_{d}\!-\!1)\left(Q(\sqrt{d}/\sigma)\!-\!\frac{1}{2}Q^{2}(\sqrt{d}/\sigma)\right)\!B(p_b, n\!-\!2d, 0, d^*\!-\!1)\!+\!
       Q(\sqrt{d}/\sigma)\!\right)
    \end{array}
    \!\!\!\!\!\right\}.
\end{eqnarray}

We have the following theorem.

\begin{theorem}
The bit-error probability can be upper-bounded by
\begin{equation}\label{BitBound}
    P_b \leq \min_{0\leq d^*\leq n}\left\{\sum\limits_{d \leq 2d^*} h'(A_d) + B(p_b, n, d^*+1, n)\right\}.
\end{equation}

\end{theorem}

\begin{IEEEproof}
Let $\hat{\underline U} \in \mathbb{F}_2^k$ be the binary output vector from a decoder
when the input to the encoder is $\underline U$. The bit-error probability associated with the decoder is defined as~\cite[p.~9]{Ryan09}
\begin{equation}\label{PbDef0}
    P_b \stackrel{\Delta}{=} \frac{1}{k} \sum_{0\leq i\leq k-1} {\rm Pr}\{\hat{u}_i \neq u_i\}.
\end{equation}
Given that the all-zero codeword is transmitted, the bit-error probability can be rewritten as
\begin{equation}\label{PbDef}
    P_b = \mathbf{E}\left\{\frac{W_H(\hat{\underline U})}{k}\right\},
\end{equation}
where $\mathbf{E}$ is the mathematical expectation.

Now we assume that Algorithm~\ref{subDEC} is implemented as the decoder.
Without loss of generality, we make an assumption that
$\hat{\underline U}$ is uniformly at random chosen from
$\mathbb{F}_2^k$ whenever Algorithm~\ref{subDEC} reports a decoding
error. Recall that $\mathcal{R} = \left\{\underline y |
{\underline c}^{(0)} \in \mathcal{L}_{\underline y}\right\}$ as defined in~(\ref{Region}).  We
assume the following partition $\mathcal{R} =
\bigcup_{d}\mathcal{R}_d$, where $\underline y \in \mathcal{R}_{d}$
if and only if Algorithm~\ref{subDEC} outputs one codeword with
Hamming weight $d$. We have
\begin{eqnarray}
  kP_b &=&  {\rm Pr}\{\underline y \in \mathcal{R}\}\mathbf{E}\{W_H(\hat{\underline U})|\underline y \in \mathcal{R}\} + {\rm Pr}\{\underline y \notin \mathcal{R}\}\mathbf{E}\{W_H(\hat{\underline U})|\underline y \notin \mathcal{R}\}\nonumber\\
  &\leq& {\rm Pr}\{\underline y \in \mathcal{R}\}\mathbf{E}\{W_H(\hat{\underline U})|\underline y \in \mathcal{R}\} + k {\rm Pr}\{\underline y \notin \mathcal{R}\}\nonumber\\
  &\leq& \sum_{d\leq 2d^*} {\rm Pr}\{\underline y \in \mathcal{R}_d\}\mathbf{E}\{W_H(\hat{\underline U})|\underline y \in \mathcal{R}_d\} + k B(p_b, n, d^*+1, n),\label{BERUnifiedBound}
\end{eqnarray}
where we have used the fact that $\mathbf{E}\{W_H(\hat{\underline
U})| \underline y \notin \mathcal{R}\} \leq k$.

Now we focus on how to upper-bound ${\rm Pr}\{\underline y \in
\mathcal{R}_d\}\mathbf{E}\{W_H(\hat{\underline U})|\underline y \in
\mathcal{R}_d\}$ for any given $d \leq 2d^*$.

On one hand,
\begin{equation}\label{BERterm1}
    \mathbf{E}\{W_H(\hat{\underline U})|\underline y
\in \mathcal{R}_d\} \leq
    \hat{i}_d
\end{equation}
by the definition of $\hat{i}_d$ and
\begin{eqnarray}\label{BERterm2}
{\rm Pr}\{\underline y \in \mathcal{R}_d\} \leq
(A_{d}-1)\left(Q(\sqrt{d}/\sigma)-\frac{1}{2}Q^{2}(\sqrt{d}/\sigma)\right)
   B(p_b, n-2d, 0, d^*-1)+Q(\sqrt{d}/\sigma)
\end{eqnarray}
from the unified upper bound (\ref{UnifiedBound}) based on
triplet-wise error probabilities.

On the other hand, we assume the following partition $\mathcal{R}_d
= \bigcup_{\ell}\mathcal{R}_d^{(\ell)}$, where $\underline y \in
\mathcal{R}_{d}^{(\ell)}$ whenever Algorithm~\ref{subDEC} outputs
${\underline c}^{(\ell)}$, $1\leq \ell \leq A_d$. Denote by
${\underline u}^{(\ell)}$ the input binary vector to the encoder
corresponding to the codeword ${\underline c}^{(\ell)}$. Since ${\rm
Pr}\{\underline y \in \mathcal{R}_d^{(\ell)}\} \leq {\rm
Pr}\{E_{0\rightarrow \ell}, \underline y \in \mathcal{R}\}$, we have
\begin{eqnarray}
    {\rm Pr}\{\underline y \in \mathcal{R}_d\}\mathbf{E}\{W_H(\hat{\underline U})|\underline y \in \mathcal{R}_d\}
    &=& \sum_{1\leq \ell\leq A_d} {\rm Pr}\{\underline y \in \mathcal{R}_d^{(\ell)}\} W_H({\underline u}^{(\ell)})\\
     &\leq& \sum_{1\leq \ell\leq A_d} {\rm Pr}\{E_{0\rightarrow \ell}, \underline y \in \mathcal{R}\} W_H({\underline u}^{(\ell)})\\
     &\leq& kA'_d Q\left(\frac{\sqrt{d}}{\sigma}\right)B(p_b, n-d, 0, d^*-1) \label{BERPairWise}
\end{eqnarray}
from the definition of $A'_d$ and
Lemma~\ref{lemma-Improved_Pair_Wise}.

Now we have two upper bounds on ${\rm Pr}\{\underline y \in
\mathcal{R}_d\}\mathbf{E}\{W_H(\hat{\underline U})|\underline y \in
\mathcal{R}_d\}$. One is~(\ref{BERPairWise}), and the other can be
obtained by combining~(\ref{BERterm1}) and (\ref{BERterm2}). Taking
into account the definition of $h'(A_d)$, we have
\begin{equation}\label{BERFirstTerm}
    {\rm Pr}\{\underline y \in \mathcal{R}_d\}\mathbf{E}\{W_H(\hat{\underline U})|\underline y \in\mathcal{R}_d\} \leq k h'(A_d).
\end{equation}
Substituting~(\ref{BERFirstTerm}) into~(\ref{BERUnifiedBound}) and
minimizing over $d^*$, we have
\begin{equation}\label{kBER}
    kP_b \leq \min\limits_{0 \leq d^* \leq n} \left\{\sum\limits_{d\leq 2d^*}  kh'(A_d) + k B(p_b, n, d^*+1, n)\right\}.
\end{equation}
Dividing by $k$ on the both sides of~(\ref{kBER}), we complete the
proof.
\end{IEEEproof}

{\bf Remark.} The bound on the bit-error probability given above is
applicable to the optimal decoding algorithm that minimizes the
bit-error probability, but will not always be applied to the ML
decoding algorithm. In other words, the ML decoding algorithm, which
is not optimal for minimizing the bit-error probability, may have a
higher bit-error probability.

\section{Numerical Results}\label{sec6}
In this section, by an $[n, k]$ {\em random linear code}, we mean a code ensemble
in which each code is defined by a uniformly at random selected full-rank parity-check matrix of size $(n-k)\times n$.
As shown in~\cite[Appendix~D]{Twitto06a}, the average weight spectra of a random linear code $[n, k]$ can be found as
\begin{equation}
 A_d = \left\{\begin{array}{rl}
                      \binom{n}{d}\frac{2^k-1}{2^n-1}, & 0 < d \leq n \\
                      1, & d=0
                    \end{array}
 \right..
\end{equation}

We also need to point out that the weight spectra of the compared BCH codes can be found in~\cite{Terada04}.

\begin{figure}
\centering
  \includegraphics[width=8cm]{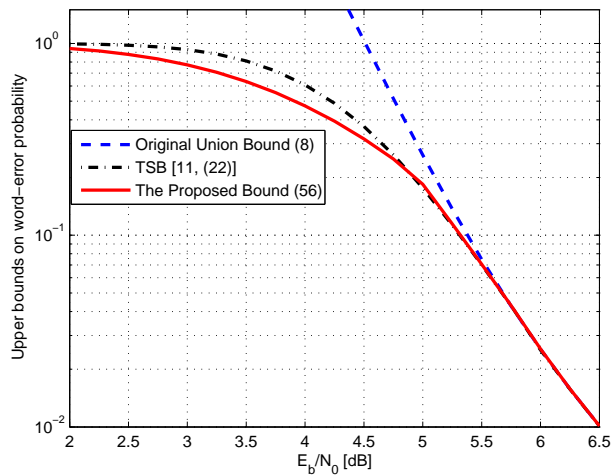}\\
  \caption{Comparison between the upper bounds on the word-error probability under ML decoding of random binary linear block codes $[100, 95]$. The compared bounds are the original union bound, the TSB and the proposed bound.}\label{proposed_union_bound_100_95}\end{figure}

\begin{figure}
\centering
  \includegraphics[width=8cm]{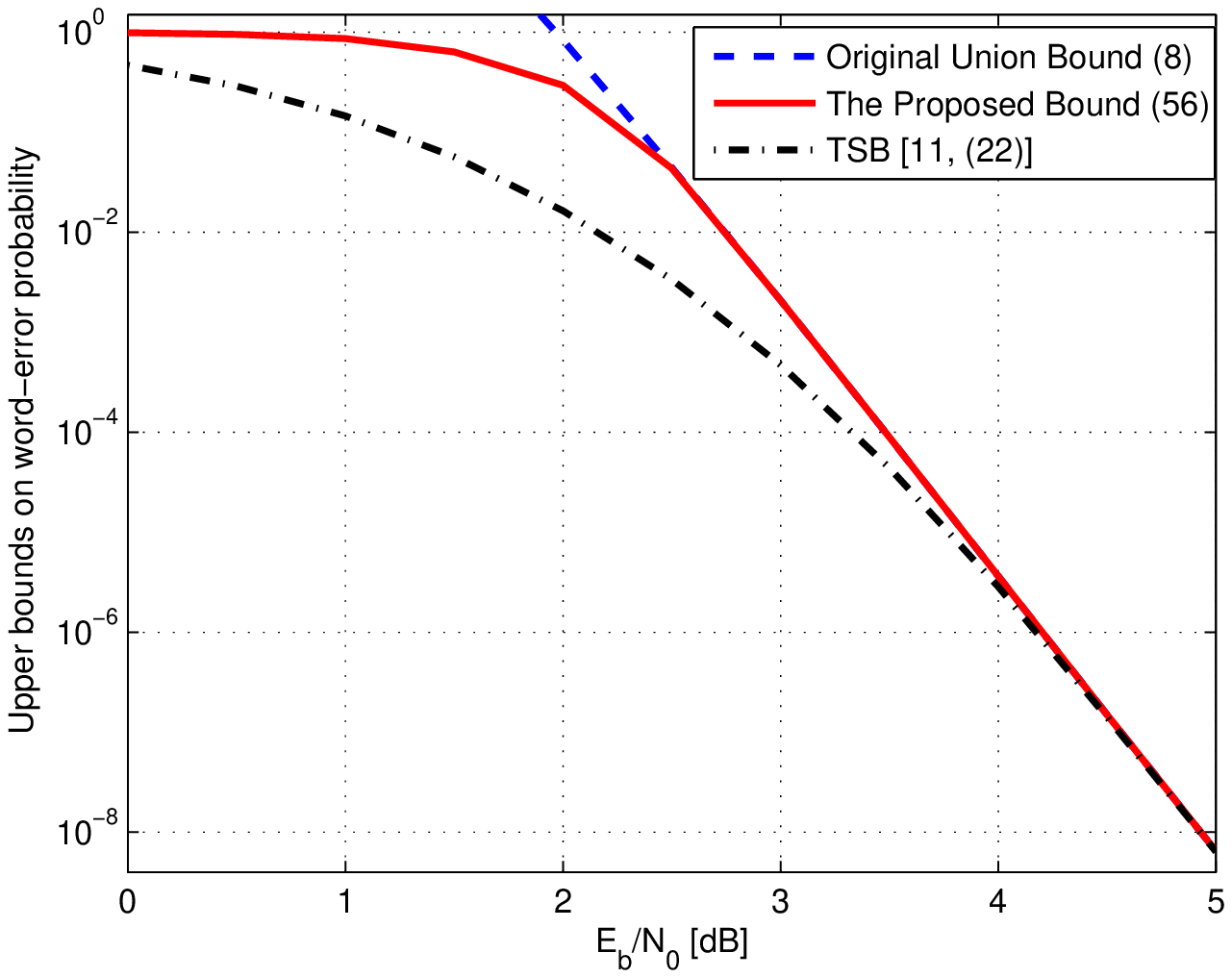}\\
  \caption{Comparison between the upper bounds on the word-error probability under ML decoding of random binary linear block codes $[100, 50]$. The compared bounds are the original union bound, the TSB and the proposed bound.}\label{proposed_union_bound_100_50}\end{figure}

\begin{figure}
\centering
  \includegraphics[width=8cm]{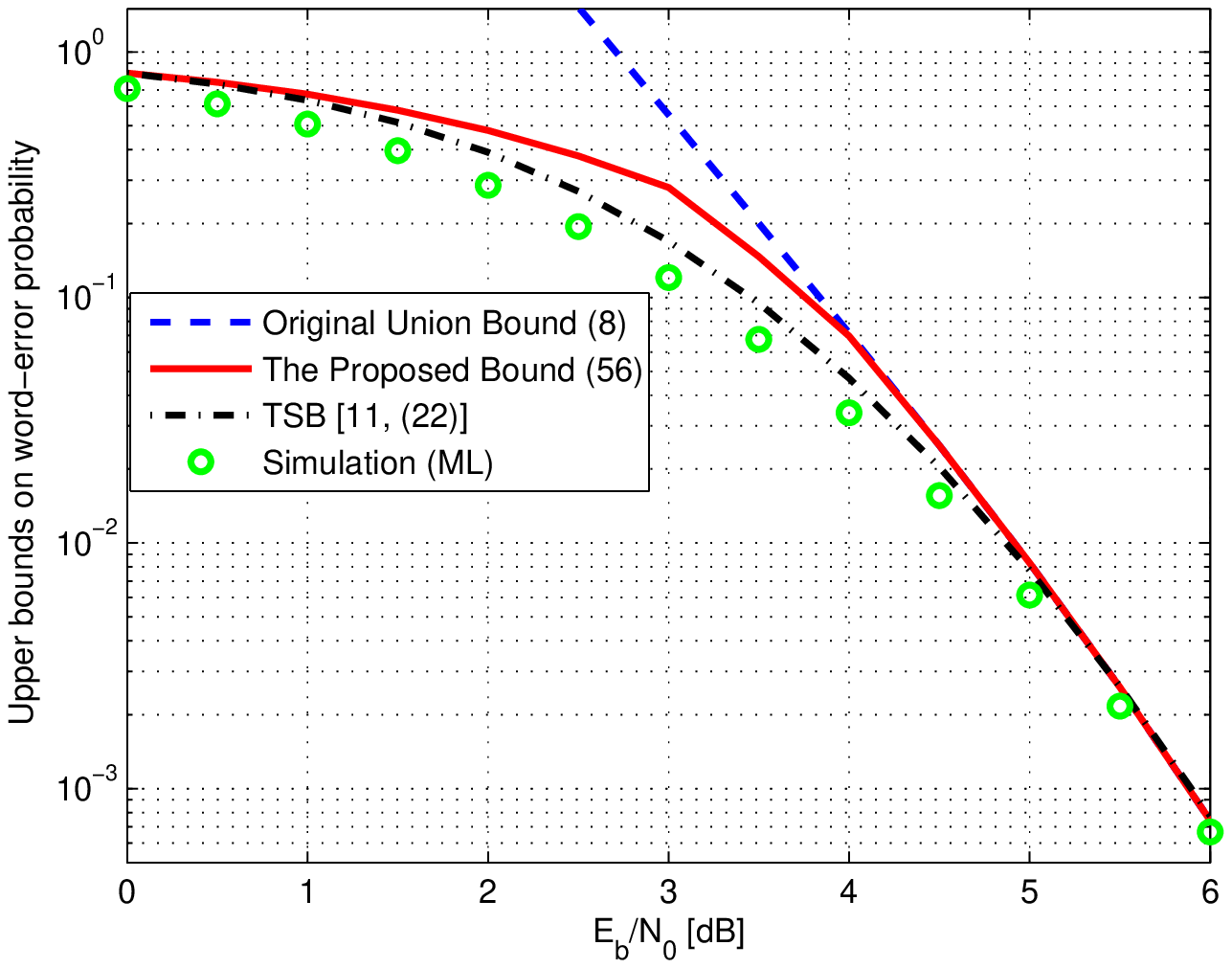}\\
  \caption{Comparison between the upper bounds on the word-error probability under ML decoding of BCH code $[31, 26]$. The compared bounds are the original union bound, the TSB and the proposed bound, which are also compared with the ML simulation results.}\label{proposed_union_bound_31_26}\end{figure}

\begin{figure}
\centering
  \includegraphics[width=8cm]{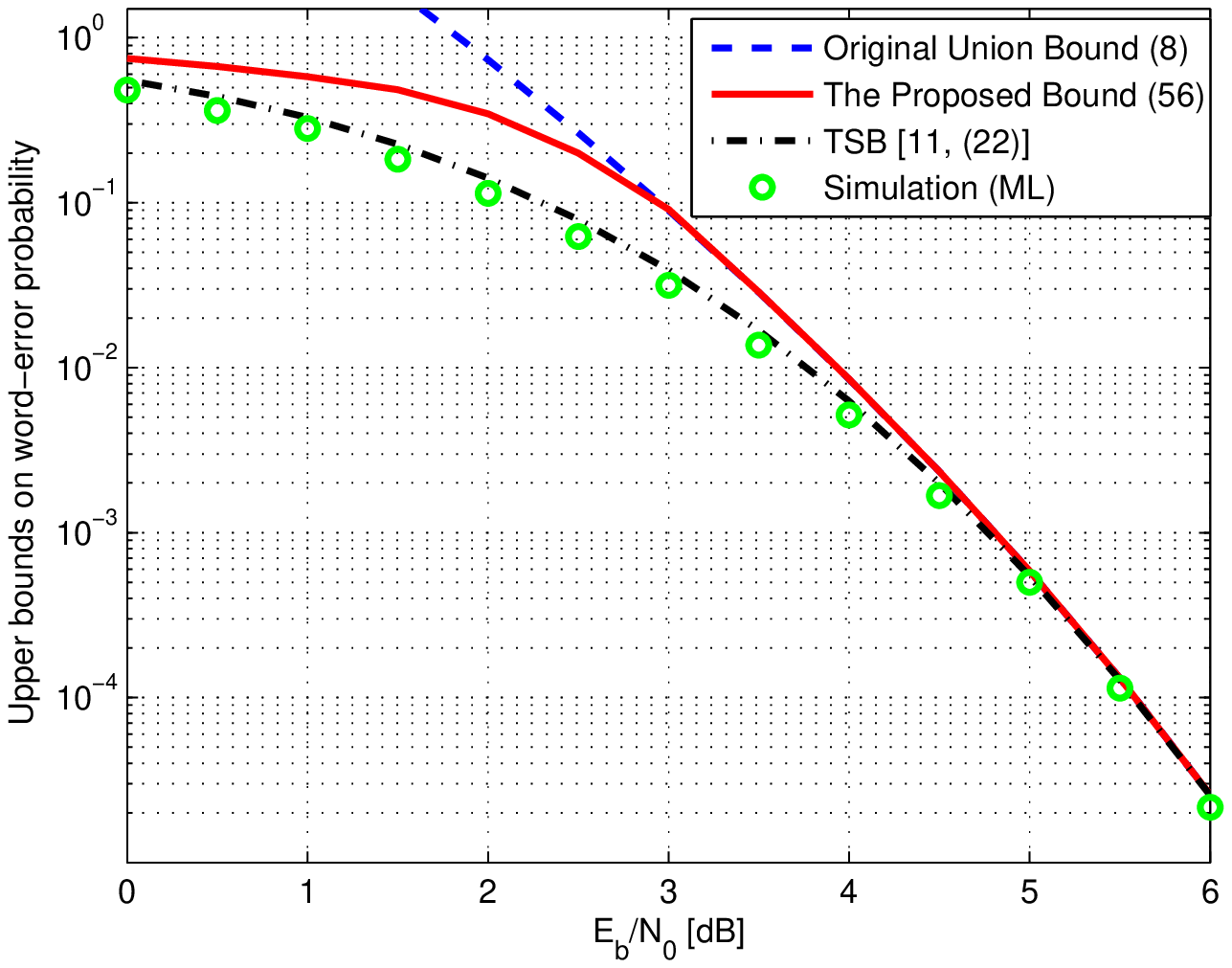}\\
  \caption{Comparison between the upper bounds on the word-error probability under ML decoding of BCH code $[31, 21]$. The compared bounds are the original union bound, the TSB and the proposed bound, which are also compared with the ML simulation results.}\label{proposed_union_bound_31_21}\end{figure}

\begin{figure}
\centering
  \includegraphics[width=8cm]{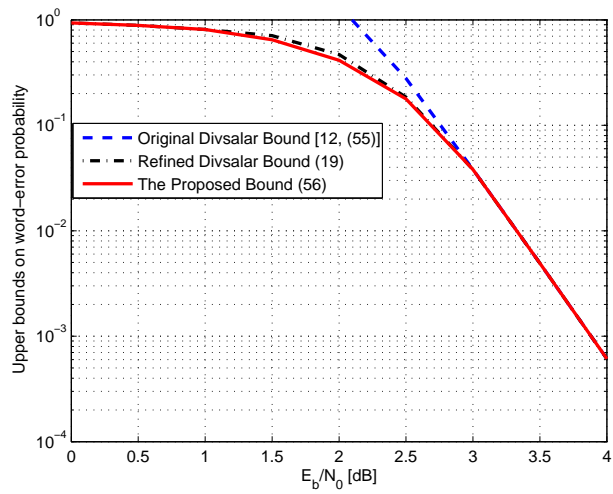}\\
  \caption{Comparison between the upper bounds on the word-error probability under ML decoding of BCH code $[63, 39]$. The compared bounds are the original Divsalar bound, the refined Divsalar bound and the proposed bound.}\label{proposed_divsalar_bound}\end{figure}

\begin{figure}
\centering
  \includegraphics[width=8cm]{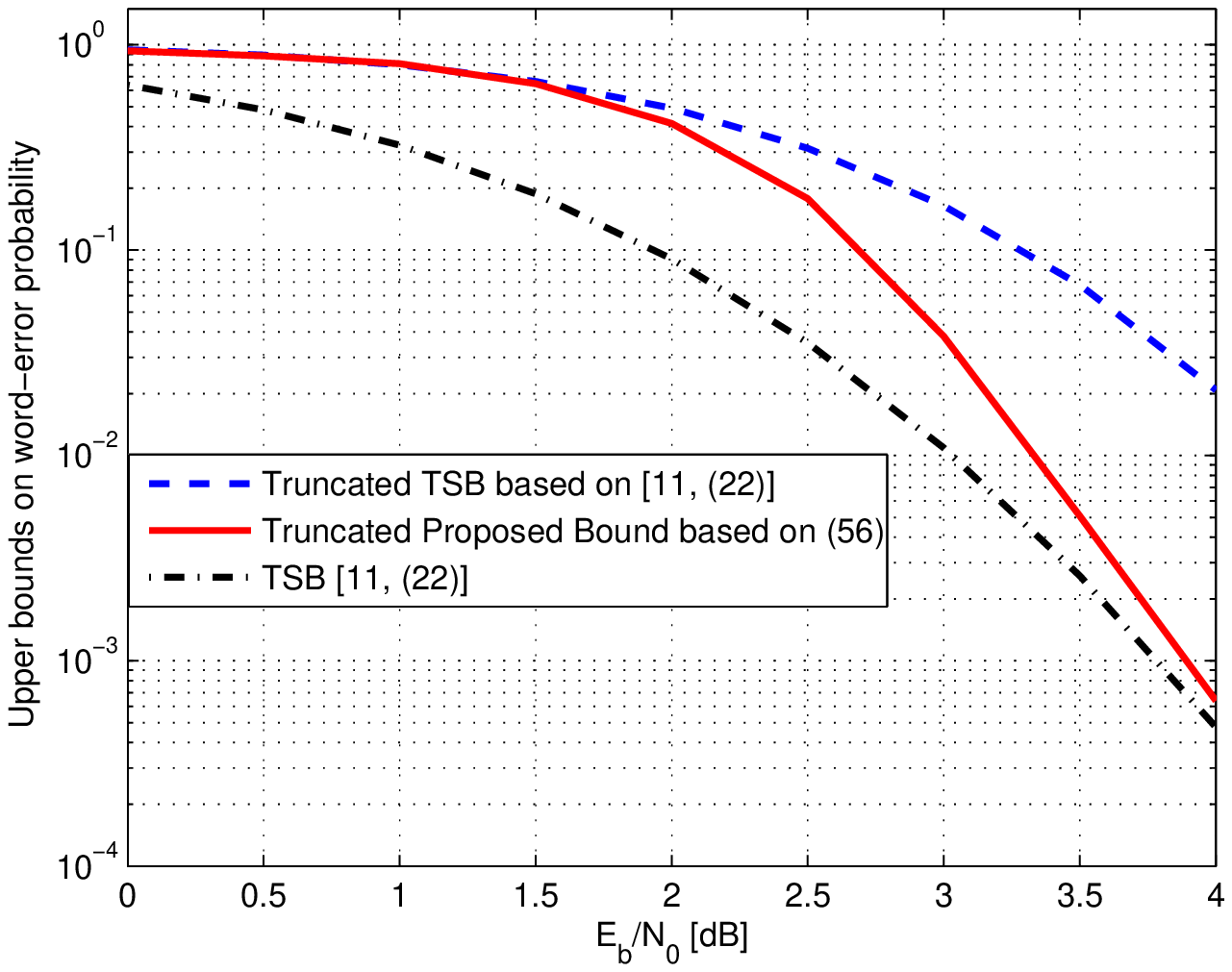}\\
  \caption{Comparison between the upper bounds on the word-error probability under ML decoding of BCH code $[63, 39]$. The compared bounds are the truncated TSB, the truncated proposed bound and the TSB. These truncated bounds depend only on the sub-code $\mathcal{C}_{20}$ consisting of all codewords with Hamming weight no greater than $20$.}\label{truncated_compare}\end{figure}

\subsection{Comparisons Between the Proposed Bounds and the Existing Bounds}

In this subsection, we present four examples to compare the proposed
bounds~(\ref{TotalBound}) with the existing bounds on word-error
probability.

Fig.~\ref{proposed_union_bound_100_95} and
Fig.~\ref{proposed_union_bound_100_50} show the comparisons between
the original union bound~(\ref{Union}), the TSB~\cite[(22)]{TSB94}
and the proposed bound~(\ref{TotalBound}) on word-error probability
of $[100, 95]$ and $[100, 50]$ random linear codes, respectively,
where the former has been used as an example in~\cite{Sason06}.
The proposed bounds are obtained by optimizing the parameter $d^{*}$,
which may be varied with SNRs. We can see that the proposed bound
improves the original union bound. We can also see that, for the
random code $[100, 95]$, the proposed bound is tighter than the TSB
in the low-SNR region; while for the random code $[100, 50]$, the
proposed bound is looser than the TSB. This coincides with the
computational results in~\cite[Fig.~3]{Twitto07a}, which tells us
that the TSB becomes looser in terms of the error exponent with
increasing code rates. Note that the solid curve in
Fig.~\ref{proposed_union_bound_100_95} is better than that
in~\cite[Fig.~3]{Ma11}, since Theorem~\ref{Theorem_UnifiedBound}
here improves~\cite[Theorem 2]{Ma11} by employing the independence
between the error events and certain components of the received
random vectors.

Fig.~\ref{proposed_union_bound_31_26} and
Fig.~\ref{proposed_union_bound_31_21} show the comparisons between
the original union bound~(\ref{Union}), the TSB~\cite[(22)]{TSB94}
and the proposed bound~(\ref{TotalBound}) on word-error probability
of $[31, 26]$ and $[31, 21]$ BCH codes, respectively. Also shown are
the simulation results. We can see that the proposed bound improves
the original union bound especially in the low-SNR region. We can
also see that the proposed bound is almost as tight as the TSB for
the $[31, 26]$ BCH code but looser than the TSB for the [31, 21] BCH
code, which again coincides with the conclusions
in~\cite{Twitto07a}.

\subsection{Combination of the Proposed Technique with the Existing Bounds}

By Proposition~\ref{proposition-optnewbound}, we know that the
proposed bounding technique can potentially improve any existing
upper bounds. To illustrate this, we give an
example. Fig.~\ref{proposed_divsalar_bound} shows the comparisons
between the original Divsalar bound~\cite[(55)]{Divsalar99}, the
{\em refined} Divsalar bound~(\ref{optnewbound}) by taking Divsalar
bound as $T_u$ and the proposed bound~(\ref{TotalBound}) on
word-error probability of $[63, 39]$ BCH code, which has been used
as an example in~\cite{TSB94}. We can see that the refined Divsalar
bound improves the original Divsalar bound especially in the low-SNR
region. We can also see that the proposed bound~(\ref{TotalBound}) is slightly tighter
than the refined Divsalar bound. For this $[63, 39]$ BCH code, we
have also combined the proposed bounding technique with the SB and
the TSB. However, we found that the optimal parameter $d^*$ is $n$
and hence no improvement is achieved for the SB and the TSB.

\subsection{Comparisons Between the Truncated Proposed Bound and the Truncated Existing Bounds}

As we have mentioned above
Proposition~\ref{proposition-optnewbound}, the proposed bounding
technique is helpful when the whole weight spectrum is unknown or
not computable, as is similar to the SB and the TSB. Hence, it makes
sense to compare these truncated bounds. To illustrate this, we take
the $[63, 39]$ BCH code as an example. To get the weight spectra,
one may need to perform the algorithms in~\cite{Barg92}. Given $d$,
the upper bounds of the computational complexity for computing $A_d$
can be found in~\cite[Lemmas 5 \& 7]{Barg92}. For example, one needs
about $10^5$ and $10^8$ attempts of Algorithm~1 in~\cite{Barg92} for
$d=9$ and $d=13$, respectively, as given in~\cite[Section
VI]{Barg92}. Evidently, the fewer $A_d$ ($0 < d \leq n$) we use, the
lower computational complexity the algorithm has. Assume that we
know only the truncated weight spectrum $\{A_d, d \leq 20\}$. Then
we can obtain the truncated proposed bound based
on~(\ref{TotalBound}) and the truncated TSB based
on~\cite[(22)]{TSB94}, as shown in Fig.~\ref{truncated_compare}.
Also shown in Fig.~\ref{truncated_compare} is the
TSB~\cite[(22)]{TSB94} with the whole weight spectrum. We can see
that the truncated proposed bound is looser than the TSB, but
tighter than the truncated TSB especially in the high-SNR region.
Note that both two truncated bounds are optimized based on the
truncated spectrum. For example, the truncated proposed bound is
obtained by optimizing the parameter $d^*$~($0\leq d^* \leq 10$)
in~(\ref{TotalBound}).

\section{Conclusions}\label{sec7}

In this paper, we have presented new techniques to improve the
conventional union bounds within the framework of GFBT. Compared with the conventional union bound, the proposed
bounds are tighter but have a similar complexity because they involve only the weight spectra and the Q-function.
The proposed bounds are also helpful when the whole weight spectrum is unknown or not
computable. Numerical results show that the proposed bounds can even improve the TSB in the high-rate region.

\appendices

\section*{Acknowledgment}
The authors would like to thank X. Huang and Q.-T. Zhuang for their
help. They also would like to thank Prof. Sason for his comments
while this work was partially presented in ISIT'2011. They also
would like to thank the Associate Editor and the anonymous reviewers
for their valuable comments.

\ifCLASSOPTIONcaptionsoff
  \newpage
\fi

% 之前用的
%\bibliographystyle{IEEEtran}
%\bibliography{tzzt}

% 秀姐用的
\small
\bibliographystyle{IEEEtranTCOM}
\bibliography{IEEEabrv,tzzt}

\end{document}